\documentclass[12pt]{article}

\usepackage[utf8]{inputenc}
\usepackage{amsmath, amssymb}
\usepackage{graphicx}
\usepackage{hyperref}
\usepackage{cite} % for compact reference formatting
\usepackage{geometry}
\title{Dereverberation Using Binary Residual Masking with Time-Domain Consistency}
\author{Daniel Williams \\ Independent Researcher}
\date{} % leave empty to omit date

\begin{document}

\maketitle

\begin{abstract}
Vocal dereverberation remains a challenging task in audio processing, particularly for real-time applications where both accuracy and efficiency are crucial. Traditional deep learning approaches often struggle to suppress reverberation without degrading vocal clarity, while recent methods that jointly predict magnitude and phase have significant computational cost. We propose a real-time dereverberation framework based on residual mask prediction in the short-time Fourier transform (STFT) domain. A U-Net architecture is trained to estimate a residual reverberation mask that suppresses late reflections while preserving direct speech components. A hybrid objective combining binary cross-entropy, residual magnitude reconstruction, and time-domain consistency further encourages both accurate suppression and perceptual quality. Together, these components enable low-latency dereverberation suitable for real-world speech and singing applications.
\end{abstract}


% ===== MAIN SECTIONS =====

\section{Introduction}
Natural reverb in vocals can degrade clarity, especially in large rooms with acoustically reflective surfaces. This can be especially harmful for people with hearing aids, teleconferencing, performances, and voice assistants, where reverb can greatly decrease the quality and intelligibility of voices. The goal of dereverberation techniques is to increase the Direct-to-Reverb Signal (DRR) ratio, reducing the reverberant tails while preserving the direct vocal signal.

In recent years, deep learning techniques have increasingly been used to remove reverb from vocal signals. Predictive masking techniques have proven especially useful for preserving vocal timbre while eliminating reverberant regions. However, typically predictive masks must alter both the phase and magnitude of the reverberant signal, forcing the model to predict separate masks for each. Some methods reuse the reverberant phase to reduce computational complexity, but since reverb is inherently temporal, this method typically decreases the perceptual accuracy of the dereverberated audio. Several methods have tried to combat this phase difference while still only predicting one mask. For instance, Zhao et al.~\cite{zhao2020} employ the Griffin-Lim phase enhancement to enhance the reverberant phase. While this technique is effective for estimating clean phase, we aim to reduce computational cost as much as possible.

In this paper, we propose a novel deep learning-based approach for speech dereverberation that employs a residual reverb mask for direct reverb tail suppression, as well as a hybrid loss function to predict a phase-aware magnitude mask for real-time dereverberation.

\section{Related Work}
Many deep learning approaches to audio processing leverage the U-Net architecture and a predictive mask to balance speed with accuracy~\cite{chung2020}. The model is trained to predict an $M \times N$ mask, where the audio file, after processed with a Short-Time Fourier Transform (STFT), is also $M \times N$. The mask can be multiplied with the reverberant audio file (after being STFT-transformed) using the Hadamard Product to yield a new clean audio segment. The benefits of using a predictive mask over full reconstruction is that the predictive mask can more easily preserve realistic features of the voice, optimally not removing any components of the direct signal.

Applying the STFT to audio files exposes both the magnitude and phase of the audio file. While dereverberating magnitude is more impactful than dereverberating phase, both have a noticeable impact in dereverberating the full vocal audio. Schwartz et al.~\cite{schwartz2024} emphasize that both magnitude and phase must be predicted to generate realistic results. They first train a model to predict the clean magnitude, using the noisy phase. Then, this output is fed into a second sub-model that predicts the real and imaginary parts of the clean signal. Zhao et al.~\cite{zhao2020} use spectral mapping on magnitude, while employing the Griffin-Lim phase enhancement to estimate the full time-domain signal. Additionally, Zhao et al. use Temporal Convolutional Networks (TCNs) with self-attention to further capture the temporal nature of reverb, giving the model temporal context in addition to magnitudinal context.

Other methods have tried to enhance both magnitude and phase without predicting two distinct multiplicative masks. Choi et al.~\cite{choi2020} propose a complex-valued predictive mask to handle both magnitude and phase, while minimizing time complexity with an optimized U-Net architecture. Williamson and Wang~\cite{williamson2017} employ a similar strategy, training the model to predict a Complex Ideal Ratio Mask (cIRM), which, when multiplied with the reverberant STFT, aims to yield the clean STFT spectrogram.

\section{Methodology}

\subsection{Signal Representation}
In order to convert the reverberant audio signal into the time-frequency domain, we apply STFT. This gives the model more context to predict a multiplicative mask to remove reverb. However, unlike traditional masking methods, our method specifically targets the reverb itself rather than the full spectrogram, predicting a residual reverb mask to remove reverb and leave the direct vocal signal untouched. The target mask is defined as:
\[
M(f,t) = \text{clip}\left(\frac{\Delta(f,t)}{R(f,t) + \epsilon}, 0, 1\right)
\]
where
\[
\Delta(f,t) = \max\{R(f,t) - C(f,t), 0\}.
\]

Here $R(f,t)$ is the magnitude of the reverberant STFT at frequency bin $f$ and time frame $t$, and $\epsilon$ is a small constant to avoid division by zero. Finally, to get the dereverberated audio, we simply apply the predicted mask to the reverberant audio and subtract the resulting spectrogram from the reverberant spectrogram:
\[
\hat{C}(f,t) = R(f,t) - \hat{\Delta}(f,t), \quad \hat{\Delta}(f,t) = M(f,t) \cdot R(f,t).
\]

The model’s input is the normalized magnitude of the reverberant signal $R(f,t)$ while the output is a single-channel magnitude mask $M(f,t)$.

\subsection{Model Architecture}
We employ a modified U-Net encoder-decoder architecture to balance low computational cost with accuracy. The encoder consists of three convolutional blocks, each followed by a max-pooling layer to downsample the feature maps. The bottleneck connects the encoder and decoder, incorporating a TCN block and a Multi-Head Attention layer. The TCN block uses a series of four dilated convolutions to increase the network’s long-term temporal dependencies. The Multi-Head Attention mechanism has four heads and a key dimension of 32, and it allows the model to weigh the importance of different time-frequency bins and capture global features.

The decoder reconstructs the feature maps back to the original input dimensions, using upsampling and skip connections from the encoder to recover details lost during downsampling. The model’s output consists of a single Conv2D layer with a $1\times1$ kernel and a sigmoid activation function. The sigmoid activation pressures the model to predict values close to either zero or one, crucial for removing all of the residual reverb while preserving all of the direct vocal signal.

\subsection{Loss Function}
We use a hybrid loss function to balance clean dereverberation with time-domain consistency. To further emphasize the goal of binary mask prediction, we use the binary cross-entropy loss between the true mask and the predicted mask as the first component of this hybrid loss function. Then, we add a weighted magnitude loss with a Gaussian emphasis on higher frequencies (close to 2 kHz) to encourage the model to apply a stronger mask where residual reverb exists, while preserving vocal timbre and clarity:
\[
L(f,t) = \text{mean}\big((M(f,t)\cdot (1-\hat{M}(f,t)))^2\big).
\]

Finally, we add a weighted time-domain loss, calculated by computing the mean-squared error between the ISTFT-predicted audio (full predicted spectrogram) and the ISTFT clean audio. We increase the weight of this loss throughout training to further encourage both frequency and time-domain dereverberation. This gives the model context about the audio signal’s phase, even though it only applies the predictive mask to the reverberant signal’s magnitude.

\subsection{Post-Processing}
After the model applies the predictive mask and subtracts the resulting spectrogram from the reverberant signal, we employ spectral gating to remove any residual reverb that the model may have missed. Additionally, we apply a simple equalizer curve to boost higher frequencies, reducing muffled effects from the prediction. Then, we stitch the audio chunks together, overlapping them with a Hann window to reduce boundary artifacts~\cite{steinmetz2023}.

\section{Experimentation}
Our dataset was synthetically generated to ensure a diverse range of training examples. Clean vocal samples were gathered from publicly available datasets, including CHiME, CommonVoice, GTSinger, The Noisy Speech Database, and the Saraga Carnatic Music Dataset, as well as our own theatre performance dataset. Reverberant/clean pairs were created by convolving each one-second audio file with a random room impulse response (RIR) from the Open AIR RIR database. We chose 13 different RIRs with long $T_{60}$ values to emphasize the residual reverb during training.

During training, we use a batch size of eight, a sample rate of 16000, a chunk size of 1638400, and a hop length of 256. We train for 125 epochs with a learning rate of 0.0001 using Adam. The best checkpoint is defined as the epoch with the lowest validation loss.

\section{Results and Evaluation}
We evaluate the model’s ability to remove reverb using the Speech-to-Reverberation Modulation Energy Ratio (SRMR), achieving an average SRMR of 3.3 in normal settings and 2.5 in heavily reverberant settings. While this is relatively low compared to state-of-the-art models, ours achieves this metric with a total latency of 9 ms (including pre- and post-processing), around 95 times faster than real-time, making it suitable for live applications.

\begin{figure}[h]
    \centering
    \includegraphics[width=0.8\linewidth]{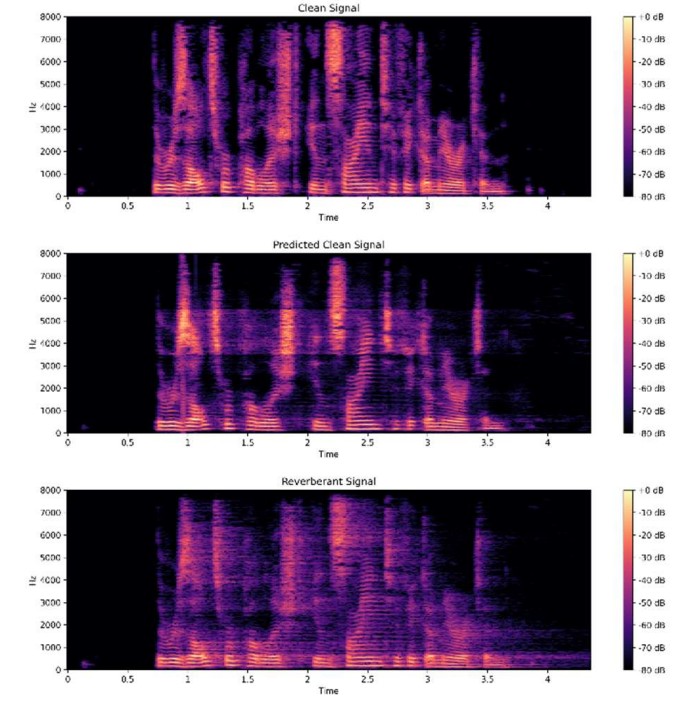}
    \caption{Comparison of clean, predicted, and reverberant spectrograms.}
    \label{fig:spectrogram}
\end{figure}

Figure~\ref{fig:spectrogram} compares spectrograms from a clean vocal signal, the reverberant input, and the dereverberated prediction. While the predicted spectrogram is less detailed than the ground truth, it preserves most of the voice while eliminating smearing and reflections.

\section{Conclusion}
Our model’s binary residual mask allows the model to target reverberant regions while leaving the original voice relatively untouched. By employing a time-domain consistency loss, the model gains crucial information about the phase, which would typically have to be predicted with a separate phase mask. This allows the model to keep a low computational cost, achieving a latency low enough for real-time applications.

However, while the model maintains very low latency, it yields dereverberated audio less detailed than the ground-truth clean audio, implying difficulty in preserving vocal timbre and clarity. Future work would aim to keep the same single-mask structure while improving the model’s ability to preserve detail in clean vocals.

% ===== REFERENCES =====


\begin{thebibliography}{9}

\bibitem{steinmetz2023}
C. J. Steinmetz, T. Walther, J. D. Reiss, ``High-Fidelity Noise Reduction With Differentiable Signal Processing,'' in \textit{Proc. AES Convention}, 2023. [Online]. Available: \url{https://www.aes.org/e-lib/browse.cfm?elib=22307}

\bibitem{chung2020}
H. Chung, V. S. Tomar, B. Champagne, ``Deep Convolutional Neural Network-Based Inverse Filtering Approach For Speech De-Reverberation,'' in \textit{Proc. IEEE MLSP}, 2020. [Online]. Available: \url{https://www.ece.mcgill.ca/~bchamp/Papers/Conference/IWMLSP2020.pdf}

\bibitem{schwartz2024}
A. Schwartz, S. Gannot, S. E. Chazan, ``Magnitude Or Phase? A Two Stage Algorithm For Dereverberation,'' in \textit{Proc. IWAENC}, 2024. [Online]. Available: \url{https://cmsworkshops.com/IWAENC2024/view_paper.php?PaperNum=1133}

\bibitem{zhao2020}
Y. Zhao, D. Wang, B. Xu, T. Zhang, ``Monaural Speech Dereverberation Using Temporal Convolutional Networks with Self-Attention,'' \textit{IEEE/ACM Trans. Audio, Speech, Lang. Process.}, 2020. [Online]. Available: \url{https://dl.acm.org/doi/abs/10.1109/TASLP.2020.2995273}

\bibitem{choi2020}
H. Choi, H. Heo, J. H. Lee, K. Lee, ``Phase-aware Single-stage Speech Denoising and Dereverberation with U-Net,'' \textit{arXiv preprint arXiv:2006.00687}, 2020. [Online]. Available: \url{https://arxiv.org/abs/2006.00687}

\bibitem{williamson2017}
D. S. Williamson, D. Wang, ``Speech Dereverberation and Denoising Using Complex Ratio Masks,'' in \textit{Proc. IEEE ICASSP}, 2017. [Online]. Available: \url{https://doi.org/10.1109/ICASSP.2017.7953226}

\end{thebibliography}
\end{document}